\RequirePackage{lineno}
\documentclass{appolb}
\usepackage{graphicx}
\usepackage{epstopdf}
\usepackage{wrapfig}

\newcommand{\pT}{$p_{\rm{\it{T}}}$~}

\newcommand{\muB}{$\mu_{\rm B}$}
\newcommand{\sNN}{$\sqrt {s}_{\rm NN}$}
\newcommand{\Npart}{$\langle N_{\mathrm{part}}  \rangle$}
\begin{document}

\title{
HIGHER MOMENTS OF THE NET-CHARGE MULTIPLICITY DISTRIBUTIONS AT RHIC ENERGIES IN STAR
}
\author{Nihar Ranjan Sahoo \\
(for the STAR Collaboration)
\address{Variable Energy Cyclotron Centre (VECC), Kolkata, India}
\\
}
\maketitle
\begin{abstract}

We report the higher order moments of the net-charge multiplicity distributions
for the Au+Au collisions at \sNN =  7.7, 11.5, 19.6, 27, 39, 62.4 and 200 GeV in the STAR experiment at the Relativistic
Heavy-Ion Collider (RHIC). The energy and centrality dependence
of higher moments and their products (such as $\sigma^2/M$, $S\sigma$ and $\kappa\sigma^{2}$) are
presented. The data are also compared to Poisson expectations and Hadron Resonance Gas model calculations.


\end{abstract}
  
\section{Introduction}
At extreme high temperature and/or chemical potential, hadronic matter undergoes a phase transition where color degrees of freedom plays important role, known as
Quark Gluon Plasma (QGP). The QCD based model calculations claim that at vanishing baryon chemical potential (\muB) the transition is simply a crossover~\cite{aoki}.
At large \muB, the phase transition is of first order~\cite{ejiri,gavai,cheng}. 
The presence of these two types of transition, at two extreme limit of  \muB,  urges to have an end point in the first order phase transition in QCD phase diagram, known as
the QCD Critical Point (CP).
To probe the CP and map the phase boundary in phase diagram, the beam energy scan program~\cite{STAR2,STAR_BES} has been undertaken with successful data taking 
in year 2010 and 2011. The STAR experiment has collected data for Au+Au collisions at \sNN = 7.7 to 200 GeV covering large range of \muB = 20 to 410 MeV ~\cite{Cleymans} in the phase diagram. \\

The various QCD based models reveal, at the CP, the susceptibilities and correlation length~($\xi$) of the system diverge~\cite{cheng}. The higher order susceptibilities and higher power of the correlation length
amplify the signal of divergence at the CP. The first four moments, such as mean ($M$), standard deviation ($\sigma$), skewness ($S$) and kurtosis ($\kappa$) of the conserved charged distributions like, net-charge, net-proton and net-strangeness are very closely related to these
respective higher order susceptibilities~\cite{cheng} and higher power of the correlation length~\cite{stephanov2009}. For example, the variance, skewness and kurtosis have been shown to be
related to the $\xi^{2}$, $\xi^{4.5}$ and $\xi^{7}$, respectively~\cite{stephanov2009}. In order to cancel the volume term in the  susceptibilities, different
combinations of the various moments are constructed, some of these are:
$ \sigma^{2}/M$,  $S\sigma$ and $\kappa\sigma^{2}$.
These products, as a function of beam energy, are expected to show
non-monotonic behavior near the CP. Recently in the STAR experiment, the analysis has been made on the higher moments of the net-charge, net-proton and net-kaon~\cite{qm12} for the beam energy scan program to probe the CP in phase diagram. 
Beside the CP search, recent lattice QCD model calculations~\cite{freezout_karsch} motivate to extract the freeze-out parameter like, temperature and chemical potential, from the experimentally measured higher moments of the net-charge multiplicity distribution.

 \section{Analysis Details}
The STAR experiment has taken data for Au+Au collisions at  \sNN = 7.7 to 200 GeV in last two years. 
 The STAR detector system provides the excellent particle identification and large acceptance for the event-by-event fluctuation analysis. 
The Time Projection Chamber (TPC) is the main tracking device. 
Extensive quality assurance is performed for each energy in order to minimize the fluctuation of detector efficiency.


 The event-by-event net-charge distribution is measured for Au +Au collisions occurring within $\pm$30 cm along the $z$ position of the interaction point from the TPC center and 2.0 cm radius in the transverse plane. The charged particles are measured between the transverse momentum (\pT) range  0.2$ < $\pT$< $2.0  GeV/{\it{c}} and pseudo-rapidity ($\eta$) range at $|\eta| < 0.5 $ region. The standard STAR track quality cuts are used for this analysis. The contamination of spallation protons, produced due to beam pipe interaction, are suppressed by removing protons within 200 $<$\pT$<$ 400 MeV/{\it{c}}. To avoid the auto-correlation effect in the higher moments analysis, the centrality selection has been done by uncorrected charged particles measured within $0.5<|\eta|< 1.0$  from the TPC detector. The heavy ion collision geometry of each event is characterized by the number of participating nucleons~($N_{\mathrm{part}}$). The average number of participant (\Npart) for each centrality bin is calculated by the Monte Carlo Glauber simulation. The finite centrality bin width effect has been avoided by using centrality bin width correction~\cite{CBW,nihar}. For the statistical error estimation, Delta theorem is used~\cite{Luo}.
\begin{figure}[tbp]
  \centering
 \includegraphics[width=0.8\textwidth]{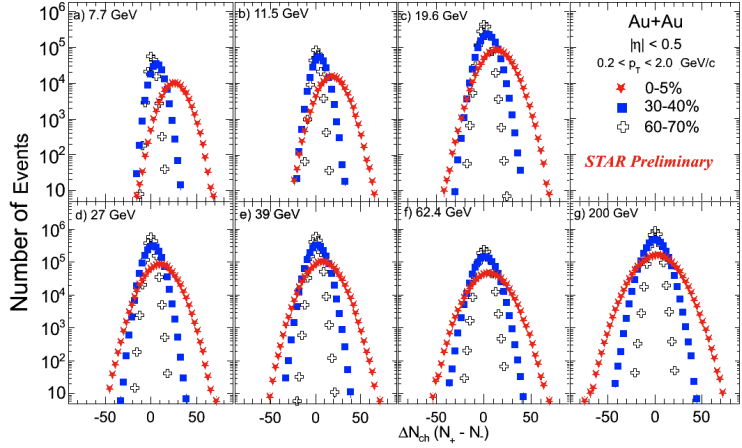} 
\caption{The raw net--charge multiplicity distributions 
 Au+Au collisions
 for three centrality bins  0-5\%, 30-40\% and 60-70\%
and collision energy for $\sqrt{s_{\rm NN}}$=7.7 to 200 GeV. }
\label{fig1}
\end{figure}

\section{Results}
\begin{figure}[ht]
\centerline{%
\includegraphics[width=9.0cm]{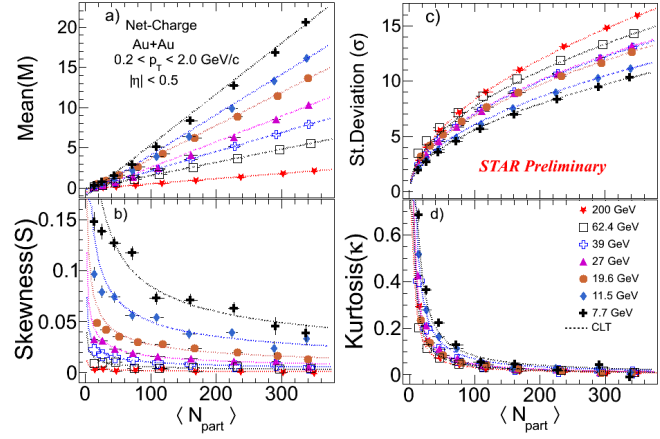}}
\caption{The various moments, mean (a), standard deviation (c), skewness (b) and kurtosis (d) are plotted with respect to \Npart ~for Au+Au collisions from \sNN=7.7 to 200 GeV. The dotted lines
represent the expectation form CLT.}
\label{fig2}
\end{figure}

The raw event-by-event net-charge multiplicity distributions are plotted in Fig.~\ref{fig1} for three centralities,  0-5\%, 30-40\% and 60-70\% for above seven colliding energies. 
It is observed that the width of net-charge multiplicity distributions increases with increase in centrality and mean of the net-charge multiplicity distributions shift towards positive side with decreasing colliding energies. In Fig.~\ref{fig2}, the moments such as, M, $\sigma$, $S$ and $\kappa$ for the net-charge distributions are plotted as function of \Npart for the above seven colliding energies. The M and $\sigma$ values increase in going from peripheral to central collisions for the above seven colliding energies, whereas $S$ and $\kappa$ values decrease with increase in collision centrality. The evolution of these higher moments can be understood by the central limit theorem (CLT)~\cite{nihar}, which explain the \Npart (proxy of volume) dependence of these moments.

\begin{figure}[ht]
\centerline{%
\includegraphics[width=9.5cm]{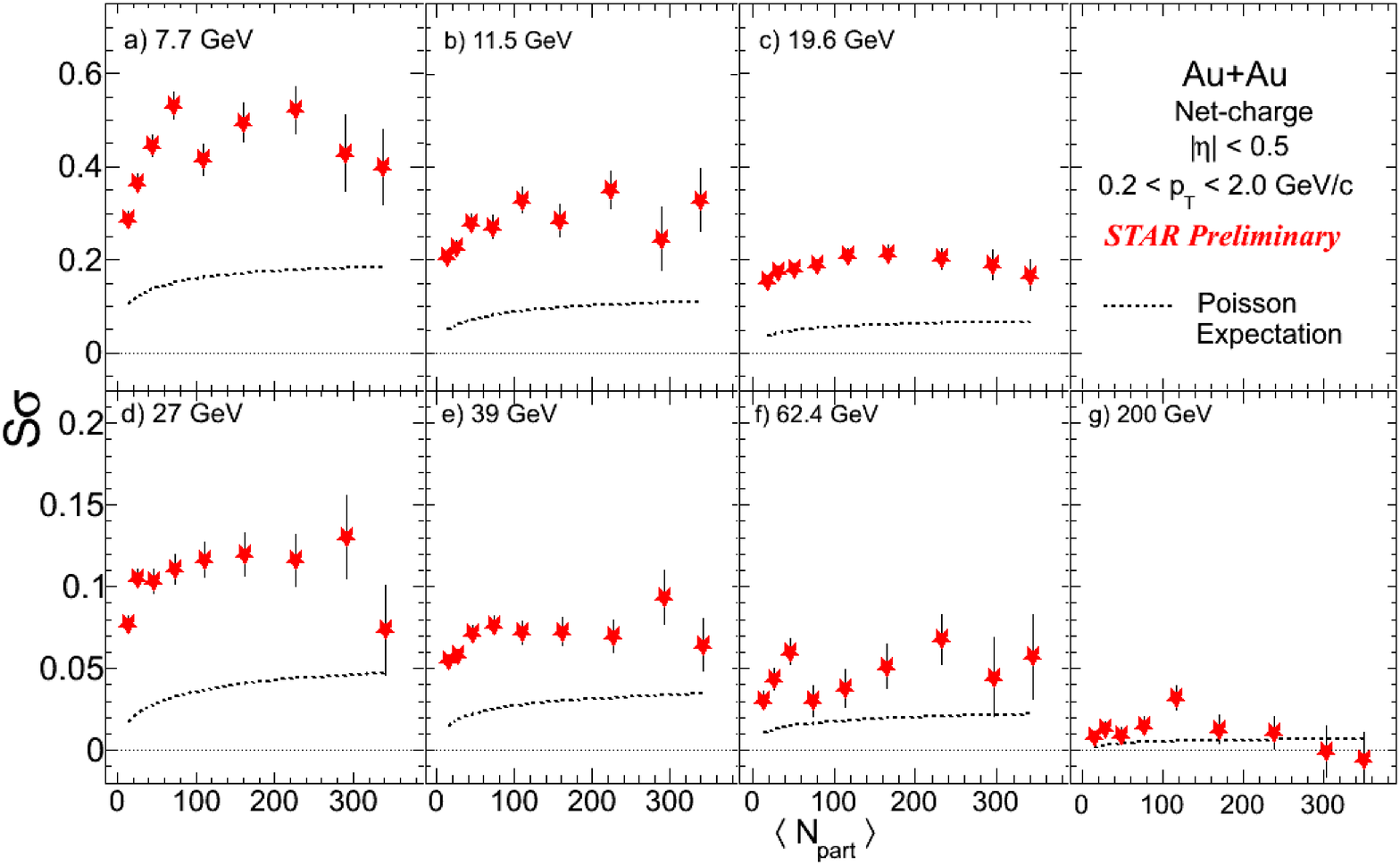}}
\caption{$S\sigma$ are plotted as a function of  \Npart ~for Au+Au collisions from \sNN=7.7 to 200 GeV. The dotted lines represent the Poisson expectation from the data.}
\label{fig3}
\end{figure}

 \begin{figure}[ht]
\centerline{%
\includegraphics[width=9.5cm]{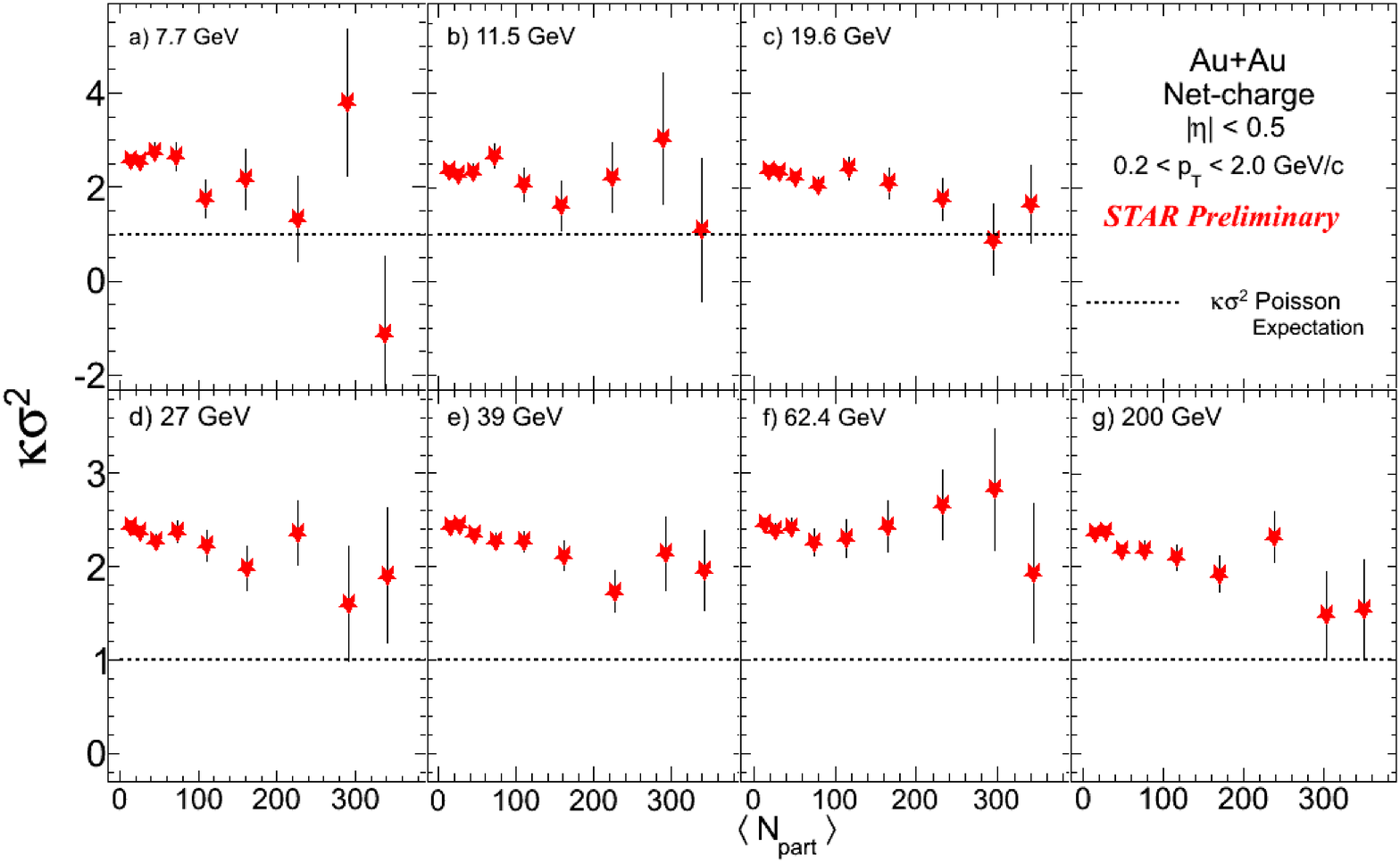}}
\caption{$\kappa\sigma^{2}$ are plotted as a function of \Npart ~for Au+Au collisions from \sNN=7.7 to 200 GeV. The dotted lines represent the Poisson expectation from the data.}
\label{fig4}
\end{figure}

  For the non-CP baseline at the above energies, the Poisson expectation is studied.
In this approach, the positive and negative charged particles distributions are assumed to be independently Poisson distribution where no dynamical correlation among the positive and negative charge particles are taken into account. Hence, the net-charge distribution is taken as a Skellam distribution (resultant of the difference of two Poisson distributions) which serve as one of the baseline for this analysis. For the Skellam distribution, $\frac{\sigma^{2}}{M} = \frac{n_{+} + n_{-}}{n_{+} - n_{-}}$, $S\sigma = \frac{n_{+} - n_{-}}{n_{+} + n_{-}}$ and $\kappa\sigma^{2} = 1$. Here $n_{+}$ and $ n_{-}$ are the means of the positive and negative charged particle distribution. The results from the Poisson expectations are discussed and compared below.

\begin{figure}[ht]
\centerline{%
\includegraphics[width=6.4cm]{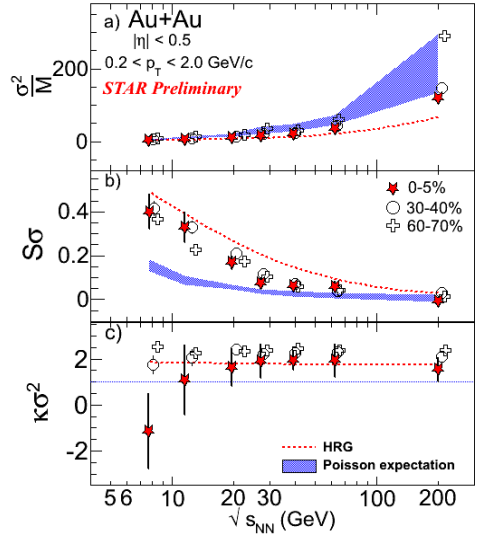}}
\caption{The $\frac{\sigma^{2}}{M}$ (a), $S\sigma$ (b) and ${\kappa\sigma^{2}}$ (c) are plotted as a function of the colliding energy for three centrality bins  0-5\%, 30-40\% and 60-70\%. The blue shaded region represent the Poisson expectation, reflects the range covering 0-5\% to 60-70\% centrality. The red dotted lines are the HRG predictions.}
\label{fig5}
\end{figure}

In Fig.~\ref{fig3} and ~\ref{fig4}, the values of $S\sigma$ and $\kappa\sigma^{2} $ are plotted as as function of \Npart for the above seven energies and corresponding Poisson expectations are also plotted for each energy. The deviation between $S\sigma$ and that of Poisson expectation are observed to be increasing as decreasing colliding energy. Whereas in $\kappa\sigma^{2}$, deviation from Poisson expectations are observed almost similar in all energies. 
 In Fig.~\ref{fig5}, the energy dependence of the product of moments such as, $\frac{\sigma^{2}}{M}$, $S\sigma$  and $\kappa\sigma^{2}$ are plotted along with Poisson expectation and Hadron Resonance Gas (HRG) model predictions. The values of $\frac{\sigma^{2}}{M}$ increases with increase in colliding energy for three centrality bins whereas in case of the $S\sigma$, it increases with decreasing colliding energies. The ${\kappa\sigma^{2}}$ shows no energy dependence and all the values are above unity, except for most central events at 7.7 GeV with large statistical uncertainty. The energy dependence of the $S\sigma$ shows systematically large deviation from its Poisson expectation below \sNN=27 GeV as compared at higher energies. The HRG predictions of ${\kappa\sigma^{2}}$ are very close to data whereas that of $S\sigma$ over-predicts the data. 
 

 \section{Summary }
 The higher moments of the net-charge multiplicity distributions have been measured for Au+Au collisions at \sNN= 7.7 to 200 GeV. The centrality dependence of the moments follows the
expectation from the CLT. The values of $\frac{\sigma^{2}}{M}$ increase with increase in colliding energy. The values of $S\sigma$ increase with decreasing colliding energies and deviates from Poisson expectation below \sNN = 27 GeV. Within statistical uncertainty, $\kappa\sigma^{2}$ is seen to be independent of collision energy.



\end{document}